\documentclass{article}
\usepackage{spconf,amsmath,graphicx}

\usepackage{caption}
\usepackage{stfloats}
\usepackage{multirow}
\usepackage{times}
\usepackage{epsfig}
\usepackage{amsmath}
\usepackage{amssymb}
\usepackage[ruled,linesnumbered]{algorithm2e}
\usepackage{graphicx}
\usepackage{algorithmic}
\usepackage{threeparttable}
\usepackage{subfigure}
\usepackage{ifpdf}

\usepackage[activate]{microtype}
\usepackage{booktabs}
\usepackage{threeparttable}

\usepackage{multicol}
\usepackage{multirow}

\usepackage{color}

\usepackage{makecell} 

\title{Improving Speaker-independent Speech Emotion Recognition Using Dynamic Joint Distribution Adaptation}
\name{Cheng Lu$^{1,2}$, Yuan Zong$^{*,1,2,4}$, Hailun Lian$^{1,3}$, Yan Zhao$^{1,3}$, Bj{\"o}rn~W.~Schuller$^5$, Wenming Zheng$^{*\thanks{$^*$Corresponding Authors.},1,2,4}$}

\address{
  $^1$Key Laboratory of Child Development and Learning Science of Ministry of Education, \\
  Southeast University, Nanjing, China\\
  $^2$School of Biological Science and Medical Engineering, Southeast University, Nanjing, China \\
  $^3$School of Information Science and Engineering, Southeast University, Nanjing, China \\
  $^4$Pazhou Lab, Guangzhou, China \\
  $^5$GLAM - the Group on Language, Audio, \& Music, Imperial College London, UK}

\begin{document}
\ninept
\maketitle
\begin{abstract}
In speaker-independent speech emotion recognition, the training and testing samples are collected from diverse speakers, leading to a multi-domain shift challenge across the feature distributions of data from different speakers. Consequently, when the trained model is confronted with data from new speakers, its performance tends to degrade. To address the issue, we propose a Dynamic Joint Distribution Adaptation (DJDA) method under the framework of multi-source domain adaptation. DJDA firstly utilizes joint distribution adaptation (JDA), involving marginal distribution adaptation (MDA) and conditional distribution adaptation (CDA), to more precisely measure the multi-domain distribution shifts caused by different speakers. This helps eliminate speaker bias in emotion features, allowing for learning discriminative and speaker-invariant speech emotion features from coarse-level to fine-level. Furthermore, we quantify the adaptation contributions of MDA and CDA within JDA by using a dynamic balance factor based on $\mathcal{A}$-Distance, promoting to effectively handle the unknown distributions encountered in data from new speakers. Experimental results demonstrate the superior performance of our DJDA as compared to other state-of-the-art (SOTA) methods.

\end{abstract}
\begin{keywords}
speaker-independent, speech emotion recognition, multi-source domain adaptation, joint distribution adaptation
\end{keywords}
\section{Introduction}

Speech emotion recognition (SER) aims to enable machines to automatically comprehend the emotions conveyed in speech signals, and it has garnered significant attention in affective computing and pattern recognition \cite{schuller2013computational}, \cite{schuller2018speech}, \cite{lu2022speech}. Particularly challenging is speaker-independent SER, which focuses on the training and speech samples collected from different speakers \cite{schuller2005speaker1}, \cite{schuller2005speaker2}.

In this case, the SER model trained on annotated speech samples from known speakers often experiences a decline in performance when dealing with new testing data from unknown speakers \cite{lu2022domain}. This degradation is primarily attributed to domain shift in feature distributions caused by speaker bias in the training data (source domain) and testing data (target domain) \cite{lu2022domain}. Fortunately, recent SER studies have demonstrated that Domain Adaptation (DA) holds promise in mitigating the domain/speaker bias in SER \cite{fan2020adaptive}, \cite{zhao22h_interspeech}, \cite{zhao2023deep}. The fundamental concept of DA is to treat the training and testing datasets collected from diverse speakers as signal source and target domains, respectively \cite{lu2022domain}. In speech representation learning, DA seeks a latent space of emotion features where the feature distributions of the source and target domains are sufficiently close. Consequently, the disparity in feature distributions betwween two domains can be eliminated, resulting in domain/speaker-invariant speech emotion features.

Although DA-based methods have shown promise in SER, they all treat the training and testing datasets as two separate domains and aim to eliminate the discrepancy across them \cite{lu2022speech}, \cite{lu2022domain}. However, since each speaker has unique pronunciation and expression habits when conveying emotions, a noticeable gap is existing in the feature distributions among different speakers' speech samples. In other words, the samples of each speaker can be considered as a separate domain.
Moreover, the speaker information of the training data is typically known in speaker-independent SER, while that of the testing data is unknown. Therefore, the feature distributions of different speakers' data exhibit a multi-source domain situation \cite{lu2022domain}. The issue of multi-domain distribution arising from different speakers poses two challenges in speaker-independent SER: (1) how to effectively and precisely measure the discrepancy across multiple domains, and (2) how to enable the model to adaptively handle unknown distributions of testing data from new speakers.

\begin{figure*}[t]
  \centering
   \includegraphics[width=5.0in,height=3.0in]{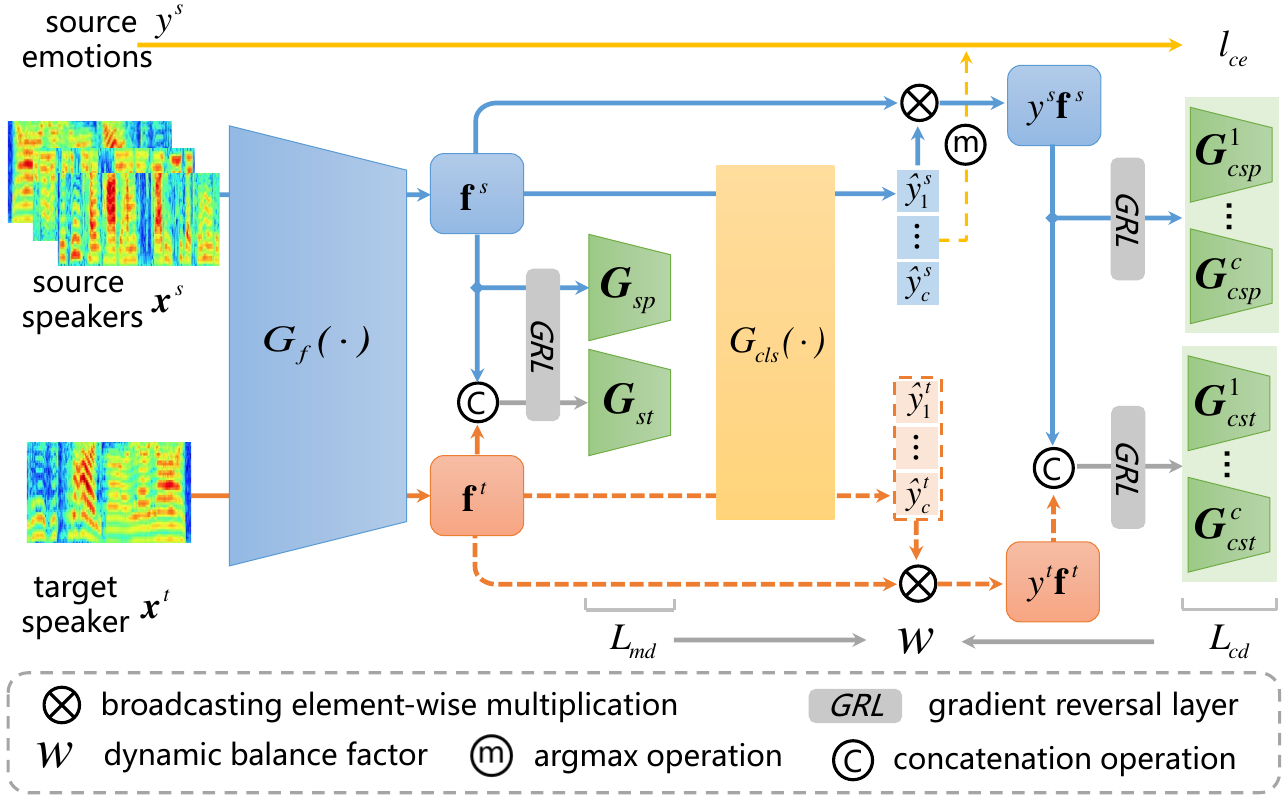}
  \caption{Overview of Dynamic Joint Distribution Adaptation (DJDA) method for speaker-independent SER, which primarily consists of JDA encompassing MDA and CDA, as well as dynamic JDA strategy.}
  \label{fig:ajda_framework}
\end{figure*}

To tackle these challenges, we propose a Dynamic Joint Distribution Adaptation (DJDA) method to address the multi-domain distribution adaptation in speaker-independent SER, shown in Figure \ref{fig:ajda_framework}. Regarding the first issue, DJDA employs a joint distribution adaptation (JDA) under multi-source domain adaptation. JDA includes both marginal distribution adaptation (MDA) and conditional distribution adaptation (CDA), which systematically measures multi-domain distribution shifts caused by speakers. JDA can achieve distribution alignment through several discriminators of domain and speaker. For the second issue, we introduce a dynamic balance factor in JDA to quantify the adaptation contributions of MDA and CDA under multi-source domain adaptation situation. This enables more flexible adaptation to the unknown distributions in data from new speakers.

\section{Proposed Method}

To facilitate the introduction, we first formalize the source and target data as ${{\mathcal{D}}_{s}}=\{(\boldsymbol{x}_{i}^{s},y_{i}^{s},y_{i}^{sp})\}_{i=1}^{{{n}_{s}}}$ and ${{\mathcal{D}}_{t}}=\{\boldsymbol{x}_{j}^{t}\}$. Herein, $\boldsymbol{x}_{i}^{s}\in {{\mathbb{R}}^{F\times T\times C}}$ and $\boldsymbol{x}_{j}^{t}\in {{\mathbb{R}}^{F\times T\times C}}$ denote the log-Mel-spectrogram features of the $i^{th}$ training and $j^{th}$ testing samples, respectively. $y_{i}^{s}\in [1,2.,..,c]$ represents the emotion label of the $i^{th}$ sample, where $c$ is the emotion number. In addition, $y_{i}^{sp}\in [1,2,...,k]$ represents the speaker label of the $k^{th}$ sample, and $k$ is the speaker number in the source dataset. $n_s$ and $n_t$ correspond to the sample numbers of source and target datasets, respectively. The input features of source and target data $\boldsymbol{x}^s$ and $\boldsymbol{x}^t$ are fed into the feature extractor $G_f(\cdot)$ to obtain their high-level emotion features ${\boldsymbol{f}}^{s}$ and ${\boldsymbol{f}}^{t}$, respectively. This process can be expressed as ${\boldsymbol{f}}^{s}={G}_{f}(\boldsymbol{x}^{s};\theta_{f})$ and ${\boldsymbol{f}}^{t}={G}_{f}(\boldsymbol{x}^{t};\theta_{f})$, where $\theta_f$ represents the parameters of $G_f(\cdot)$. Then, we will construct Joint Distribution Adaptation (JDA) through marginal distribution adaptation (MDA) and conditional distribution adaptation (CDA).

\subsection{Joint Distribution Adaptation Under Multi-source Domain Adaptation}


\subsubsection{Marginal Distribution Adaptation (MDA)}
MDA aims to address the distribution discrepancy between the source and target data, as well as among multiple speakers within the source data. To accomplish these goals, we design the source-target domain discriminator $G_{st}(\boldsymbol{f}^{s},\boldsymbol{f}^{t};\theta_{st})$ and the source-domain speaker discriminator $G_{sp}(\boldsymbol{f}^{s};\theta_{sp})$. Here, $\theta_{st}$ and $\theta_{sp}$ represent the network parameters of the discriminators $G_{st}(\cdot)$ and $G_{sp}(\cdot)$, respectively. Through an adversarial training strategy to update the parameters $\theta_{st}$ and $\theta_{sp}$, $G_{st}(\cdot)$ and $G_{sp}(\cdot)$ can effectively obscure domain information within speech representations, resulting in the generation of domain/speaker invariant features. Thus, MDA loss $\mathcal{L}_{md}$ can be obtained by combining the source-target released MDA loss $\mathcal{L}_{st}$ from $G_{st}(\cdot)$ and the speaker-related MDA loss $\mathcal{L}_{sp}$ from $G_{sp}(\cdot)$:
\begin{eqnarray}
\begin{aligned}
& \mathcal{L}_{md}\left(\theta_f, \theta_{st}, \theta_{s p}\right)
 =\mathcal{L}_{st}\left(\theta_f, \theta_{st}\right)+\mathcal{L}_{sp}\left(\theta_f, \theta_{sp}\right) \\
& =\frac{1}{n_s+n_t} \sum_{\substack{\boldsymbol{x}_i \in \\ \left(\mathcal{D}_s \cup \mathcal{D}_t\right)}} J\left(G_{st}\left(G_f\left(\boldsymbol{x}_i\right)\right), d_i\right) \\
& +\frac{1}{n_s} \sum_{\boldsymbol{x}_i \in D_s} J\left(G_{sp}\left(G_f\left(\boldsymbol{x}_i^s\right)\right), y_i^{sp}\right),
\end{aligned}
\label{e1}
\end{eqnarray}
where $d_i\in \{0,1\}$ represents the domain label of the $i^{th}$ speech sample, and the domain labels of the source and target domains are $0$ and $1$, respectively. $y_{i}^{sp}\in [1,2,...,k]$ denotes the speaker label of the $i^{th}$ speech sample. $J(\cdot)$ is the cross-entropy loss function.

\subsubsection{Conditional Distribution Adaptation (CDA)}

MDA focuses on the global distribution adaptation across multiple domains, while ignoring the local multi-class structure of emotional features. Therefore, we consider class-wise CDA to measure the fine-grained discrepancy in local feature distributions. To achieve this goal, we also need to confuse the information both between source and target domains as well as inside speaker domains under each class's features.

Similar to MDA, we first design a set of class-wise domain discriminators $\{G^{m}_{cst}\}_{m=1}^{c}$ to confuse source-target domain information of each class features, where $G^{m}_{cst}$ represents the domain discriminator of the $m^{th}$ class's features. Furthermore, we note that each $i^{th}$ sample can generate emotion prediction probabilities through the emotion classifier $G_{cls}(\cdot)$ with its parameters $\theta_y$, denoted as $\hat{y}i=G_{cls}(\boldsymbol{f}^s,\boldsymbol{f}^t;\theta_y)$. The probabilities provide valuable supervision for assessing the classification of each emotion. Consequently, we assign these probabilities weights to their corresponding features for enabling the domain discriminator to accurately distinguish domain labels bound on correct emotion prediction, thereby facilitating subsequent domain confusion. Through this process, we obtain the domain-related CDA loss function ${{\mathcal{L}}_{dcd}}$ from the class-wise source-target domain discriminator:
\begin{equation}
\begin{aligned}
& \mathcal{L}_{dcd}\left(\theta_f,\left.\theta_{cst}^m\right|_{m=1} ^c\right)\\
& =\frac{1}{n_s+n_t} \sum_{m=1}^c \sum_{\substack{\boldsymbol{x}_i \in \\ \left(\mathcal{D}_s \cup \mathcal{D}_t\right)}} \mathcal{L}_{cst}^m\left(G_{cst}^m\left(\hat{y}_i^m G_f\left(\boldsymbol{x}_i\right)\right), d_i\right),
\end{aligned}
\label{e2}
\end{equation}
where $\hat{y}_{i}^{m}$ represents the predicted probability of the $m^{th}$ class of the $i^{th}$ sample. $\mathcal{L}_{cst}^{m}$ is the source-target domain discriminator loss for the $m^{th}$ class features calculated by the cross-entropy loss function.

Note that $G_{cls}(\cdot)$ is trained based on source data, meaning that the emotion labels obtained for target data are pseudo-labels. As the network parameters are continuously optimized, the pseudo-labels will be gradually improved.

Then, we construct a set of class-wise speaker discriminators $\{G_{csp}^{m}\}_{m=1}^{c}$ to discriminate the speaker labels of each emotion features in the source domain, where $G_{csp}^{m}$ denotes the speaker discriminator for the $m^{th}$ class's features.
Similarly, the predicted probability of each class is also utilized to weight the corresponding class features for ensuring that the speaker domain labels can be more accurately distinguished bound on the correct predicted emotions, which provides the basis for speaker information confusion. Therefore, the speaker-related CDA loss function ${{\mathcal{L}}_{scd}}$ from the class-wise speaker discriminator can be denoted as follows:
\begin{equation}
\begin{aligned}
& \mathcal{L}_{scd}\left(\theta_f, \theta_y,\left.\theta_{csp}^m\right|_{m=1} ^c\right)\\
& =\frac{1}{n_s} \sum_{m=1}^c \sum_{\boldsymbol{x}_i \in \mathcal{D}_s} \mathcal{L}_{csp}^m\left(G_{csp}^m\left(\hat{y}_i^m G_f\left(\boldsymbol{x}_i\right)\right), y_i^{sp}\right).
\end{aligned}
\label{e3}
\end{equation}

Consequently, we combine the Equation \ref{e2} and \ref{e3} to produce the overall CDA loss ${{\mathcal{L}}_{cd}}$ as follows:
\begin{equation}
\begin{aligned}
& \mathcal{L}_{c d}\left(\theta_f, \theta_y,\left.\left[\theta_{c s t}^m, \theta_{c s p}^m\right]\right|_{m=1} ^c\right)\\
& =\mathcal{L}_{dcd}\left(\theta_f,\left.\theta_{cst}^m\right|_{m=1} ^c\right) +\mathcal{L}_{scd}\left(\theta_f, \theta_y,\left.\theta_{csp}^m\right|_{m=1} ^c\right).
\end{aligned}
\label{e4}
\end{equation}

\subsection{Dynamic Joint Distribution Adaptation Under Multi-source Domain Adaptation}

Due to the unknown and diverse data distributions in the target domain, a practical approach is to empirically set a balance coefficient to adjust the contributions of MDA and CDA for JDA. However, this strategy lacks flexibility. Wang et al.\ \cite{yu2019transfer}, \cite{wang2020transfer} firstly introduced a dynamic adversarial factor based on $\mathcal{A}$-distance to balance different distribution adaptation for signal-source DA. According to \cite{ben2006analysis}, the $\mathcal{A}$-distance is related to minimum errors $err(h)$ of classifiers (e.\,g., linear classifiers) when distinguishing the source and target domains, which can be defined as ${{d}_{A}}({{\mathcal{D}}{s}},{{\mathcal{D}}_{t}})=2(1-2err(h))$.

Inspired by \cite{yu2019transfer}, \cite{wang2020transfer}, we propose a dynamic balance factor to achieve adaptive JDA under multi-source domain adaptation. To produce the dynamic factor, we first calculate the $\mathcal{A}$-distance of the multi-source domain marginal distribution. According to the definition of $\mathcal{A}$-distance, we can utilize the MDA loss of the source-target domain and the speaker sub-domains of the source data obtained by multiple discriminators as the unbiased estimator of $\mathcal{A}$-distance. Therefore, we define the $\mathcal{A}$-distance of the marginal distribution as ${{d}_{md}}=2(1-2{{\mathcal{L}}_{m}})$. Similarly, the $\mathcal{A}$-distance of the conditional distribution under $m^{th}$ emotion class ${{d}_{cd}^m}$ can be denoted as ${{d}_{cd}^m}=2(1-2(\mathcal{L}_{cst}^{m}+\mathcal{L}_{csp}^{m}))$.

Combining the $\mathcal{A}$-distance of MDA and CDA, the dynamic balance factor $w$ of JDA under the multi-source DA can be generated as follows:
\begin{equation}
\begin{aligned}
w=1-\frac{{{d}_{md}}}{{{d}_{md}}+\sum\nolimits_{m=1}^{c}{{{d}^{m}_{cd}}}}.
\end{aligned}
\end{equation}

Finally, we can obtain the total loss $\mathcal{L}_{total}$ for DJDA under multi-source domain adaptation as follows:
\begin{equation}
\begin{aligned}
\mathcal{L}_{total}=\mathcal{L}_{ce}-\eta \cdot\left((1-w) \cdot \mathcal{L}_{md}+w \cdot \mathcal{L}_{cd}\right), \\
\end{aligned}
\label{e5}
\end{equation}
where $\mathcal{L}_{c e}\left(\theta_f, \theta_y\right)=\sum_{\boldsymbol{x}_i \in \mathcal{D}_s} J\left(G_f\left(\boldsymbol{x}_i^s\right), y_i^s\right)$ represents the cross-entropy loss function of the emotion class, and $\eta $ is the penalty coefficient to balance different items.

\section{Experiments}

\subsection{Experimental Dataset}
The experiments utilize two public speech emotion datasets: the English Multimodal Emotion Database (IEMOCAP) \cite{busso2008iemocap} and the Berlin German Emotion Dataset (Emo-DB) \cite{burkhardt2005database}. In detail, we utilize the improvised data in the IEMOCAP dataset, comprising a total of 2280 speech samples with 4 emotions: Angry, Happy, Sad, and Neutral. For the Emo-DB, 535 speech samples consisting of 7 emotions: Anger, Boredom, Disgust, Fear, Happiness, Sadness, and Neutral, are adopted to conduct our experiments.

\subsection{Experimental Setting}
In our experiments, the log-Mel-spectrograms of all speech samples are extracted as the input features. Moreover, to evaluate the method's performance effectively, the leave-one-speaker-out (LOSO) cross-validation \cite{schuller2009acoustic}~\cite{stuhlsatz2011deep} is adopted as the experimental protocol. In terms of evaluation metrics, we employ Weighted Average Recall (WAR) and Unweighted Average Recall (UAR) to assess the recognition accuracy of comparison methods \cite{schuller2009acoustic}.

For the proposed DJDA, We choose VGGNet as the feature extractor \cite{lu2022domain}. The source-target and speaker discriminators in MDA are implemented as two-layer Multilayer Perception (MLP) with dimensions of (1024, 512, 2) and (1024, 512, speaker number), respectively. These discriminators resemble those used in CDA. Our DJDA model is implemented using PyTorch on NVIDIA GTX3090 GPUs and optimized using the Adam optimizer with an initial learning rate of 0.0005 and a batch size of 32.

\begin{table}[t]
	\caption{Experimental results of different comparison methods on IEMOCAP, where the bold represents the best result.}
	\centering
	\begin{tabular}{|c|c|cc|}
		\hline
		\multirow{2}{*}{Protocol}   & \multirow{2}{*}{Comparison Method} & \multicolumn{2}{c|}{Accuracy (\%)}       \\ \cline{3-4}
		&                                    & \multicolumn{1}{c|}{WAR}   & UAR   \\ \hline \hline
		\multirow{9}{*}{\begin{tabular}[c]{@{}c@{}}LOSO \\ (10 speakers \\ or \\ 5 sessions) \end{tabular}}
        & DNN-HMM \cite{mao2019revisiting}      & \multicolumn{1}{c|}{62.28} & 58.02 \\ \cline{2-4}
		& CNN+LSTM Model \cite{xie2019speech}           & \multicolumn{1}{c|}{68.80} & 59.40 \\ \cline{2-4}
		& CNN GRU-SeqCap \cite{wu2019speech}           & \multicolumn{1}{c|}{72.73} & 59.71 \\ \cline{2-4}
		& FCN+Attention \cite{zhang2018attention}      & \multicolumn{1}{c|}{70.40} & 63.90 \\ \cline{2-4}
		& Model-3 Fusion \cite{bhosale2020deep}          & \multicolumn{1}{c|}{72.34} & 58.31 \\ \cline{2-4}
		& ADARL \cite{fan2020adaptive}                    & \multicolumn{1}{c|}{73.02} & 65.86 \\ \cline{2-4}
		
		& ATFNN  \cite{lu2022speech}                & \multicolumn{1}{c|}{73.81} & 64.48 \\ \cline{2-4}  \cline{2-4}
		& DJDA (ours)                               & \multicolumn{1}{c|}{\textbf{75.26}} & \textbf{65.92} \\ \hline
	\end{tabular}
	\label{tab:nb1}
\end{table}

\begin{table}[t]
	\caption{Experimental results of different comparison methods on Emo-DB, where the bold represents the best result.}
	\centering
	\begin{tabular}{|c|c|cc|}
		\hline
		\multirow{2}{*}{Protocol}  & \multirow{2}{*}{Comparison Method} & \multicolumn{2}{c|}{Accuracy (\%)}                            \\ \cline{3-4}
		&                                    & \multicolumn{1}{c|}{WAR}   & UAR                        \\ \hline\hline
		\multirow{11}{*}{\begin{tabular}[c]{@{}c@{}}LOSO \\ (10 speakers)   \end{tabular}}
		& GerDA \cite{stuhlsatz2011deep}                     & \multicolumn{1}{c|}{81.90} & 79.10                      \\ \cline{2-4}
		& DNN\_ELM \cite{han2014speech}                 & \multicolumn{1}{c|}{77.01} & 76.98                      \\ \cline{2-4}
		& ComParE\_SVM \cite{eyben2015geneva}             & \multicolumn{1}{c|}{N/A}   & 86.00                      \\ \cline{2-4}
		& SDFA \cite{zhao2019speech}                      & \multicolumn{1}{c|}{86.65} & N/A                        \\ \cline{2-4}
		& DTPM \cite{zhang2017speech}                     & \multicolumn{1}{c|}{87.31} & 86.30                      \\ \cline{2-4}
		& DANN \cite{tu2019towards}                       & \multicolumn{1}{c|}{85.98} & 84.61                      \\ \cline{2-4}
		& DAN \cite{long2015learning}                     & \multicolumn{1}{c|}{86.36} & 85.30                      \\ \cline{2-4}
		& DIFL\_VGG6 \cite{lu2022domain}                  &  \multicolumn{1}{c|}{89.72}  &  88.49  \\ \cline{2-4} \cline{2-4}
		& DJDA (ours)                               &  \multicolumn{1}{c|}{\textbf{89.91}}  &  \textbf{88.69} \\ \hline
	\end{tabular}
	\label{tab:nb2}
\end{table}

\subsection{Results and Analysis}
For the experiments on IEMOCAP, we selected several state-of-the-art (SOTA) models as comparison methods, i.\,e., DNN-HMM \cite{mao2019revisiting}, CNN+LSTM Model \cite{xie2019speech}, CNN\_GRU-SeqCap \cite{wu2019speech}, FCN$+$Attention \cite{zhang2018attention}, Model-3 fusion \cite{bhosale2020deep}, ADARL \cite{fan2020adaptive}, and ATFNN \cite{lu2022speech}. The comparison results are presented in Table~\ref{tab:nb1}. These results indicate that DJDA achieves the highest accuracy in both WAR (75.26\%) and UAR (65.92\%). Specifically, DJDA outperforms deep learning-based baseline methods (e.\,g., DNN-HMM, CNN+LSTM Model, and CNN\_GRU-SeqCap) by more than 5\% in WAR results, and its performance also surpasses DA-based methods (e.\,g., ADARL).

In the experiments on Emo-DB, we chose GerDA~\cite{stuhlsatz2011deep}, DNN\_ELM \cite{han2014speech}, ComParE\_SVM \cite{eyben2015geneva}, SDFA \cite{zhao2019speech}, DTPM \cite{zhang2017speech}, DANN \cite{tu2019towards}, DAN \cite{long2015learning}, and DIFL\_VGG6 \cite{lu2022domain} as benchmark experimental methods. The comparison results of different methods are shown in Table~\ref{tab:nb2}, where it is evident that DJDA achieves the best recognition performance, with a WAR of 89.91\% and UAR of 88.69\%. Furthermore, the results in Table~\ref{tab:nb2} also demonstrate that DA-based methods (e.\,g., DANN, DAN, and DIFL\_VGG6) generally attain higher accuracy compared to non-DA methods (e.\,g., GerDA, DNN\_ELM, and DTPM). This can be attributed to the fact that DANN, DAN, and DIFL\_VGG6 consider the domain shifts caused by diverse speakers between training and testing data, thereby obtaining more speaker-invariant emotion features than non-DA-based methods. Moreover, our proposed DJDA exhibits better recognition performance than the DA-based method, indicating that dynamic joint distribution adaptation is more effective than single marginal distribution adaptation or conditional distribution adaptation.

Consequently, it is evident that our DJDA achieves SOTA performance on IEMOCAP and Emo-DB, thereby fully demonstrating the effectiveness of DJDA for addressing speaker-independent SER.

\begin{figure}[ht]
  \centering
  \includegraphics[width=0.9\linewidth]{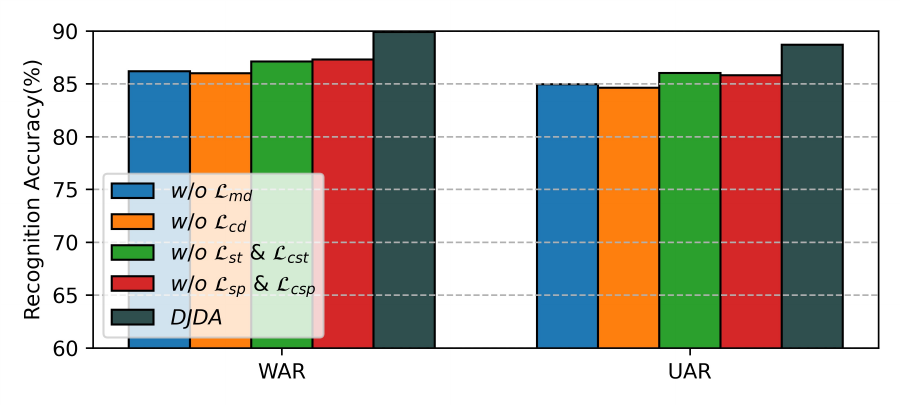}
  \caption{Ablation experiments of DJDA on Emo-DB.}
  \label{fig:ablation}
\end{figure}

\subsection{Ablation Experiments}
To verify the performance of different modules in DJDA, we also design the corresponding ablation experiments on Emo-DB, and experimental results are presented in Figure~\ref{fig:ablation}. Among them, w/o $\mathcal{L}_{md}$, w/o $\mathcal{L}_{cd}$, w/o $\mathcal{L}_{st} \& \mathcal{L}_{cst}$, and w/o $\mathcal{L}_{sp} \& \mathcal{L}_{csp}$ represent DJDA only with CDA, DJDA only with MDA, DJDA only with speaker-wise JDA, and DJDA only with source-target-wise JDA, respectively.

The experimental results in Figure~\ref{fig:ablation} show that our proposed DJDA obtains the optimal results, indicating that DJDA can indeed effectively preserve the speaker-invariant robustness of speech emotion features. Specifically, CDA obtains better results than MDA due to its function on distribution discrepancy elimination of finer-grained various emotional features. Furthermore, the results of JDA are better than MDA and CDA because of similar reasons. The speaker-wise domain adaptation (w/o $\mathcal{L}_{st} \& \mathcal{L}_{cst}$) achieves comparable performance to the source-target-wise domain adaptation (w/o $\mathcal{L}_{sp} \& \mathcal{L}_{csp}$).

\subsection{Discussion on the dynamic balance coefficient}
\begin{figure}[t]
\centering
\subfigure[$w$ in the IEMOCAP database]{
\label{iemocap-cm}
\includegraphics[width=0.48\linewidth]{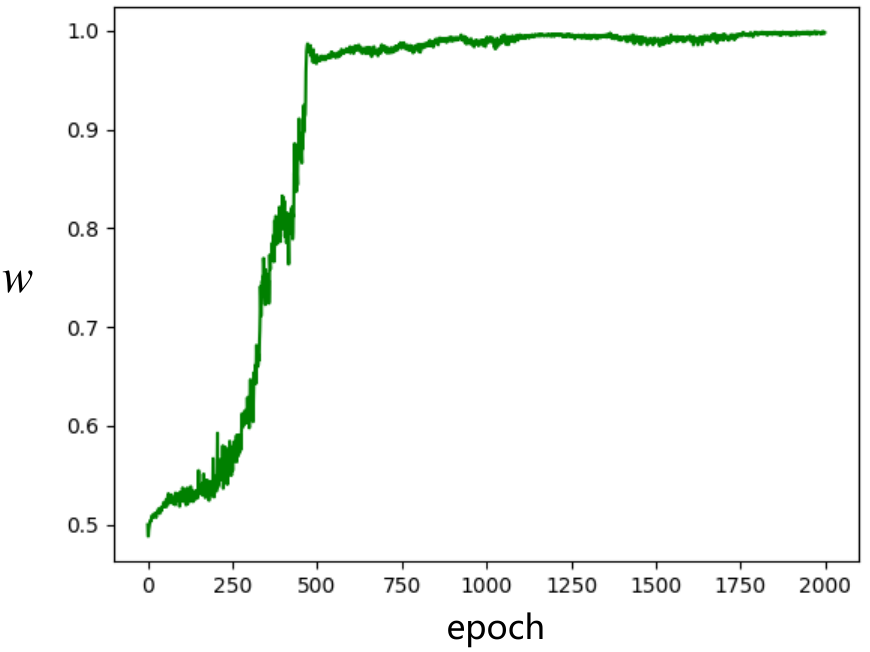}}
\subfigure[$w$ in the Emo-DB database]{
\label{abc-cm}
\includegraphics[width=0.48\linewidth]{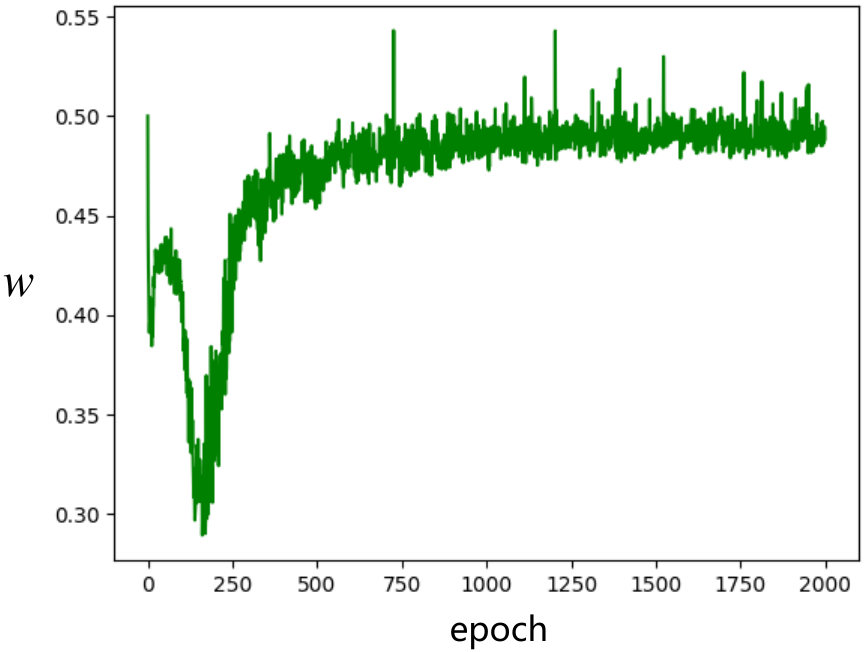}}

\caption{Dynamic balance factors for different target samples. }
\label{w}
\end{figure}

This section aims to investigate the performance of the dynamic balance coefficient in DJDA when applied to unknown distributions from different speakers. We visualize the variation of the dynamic balance coefficient values $w$ on IEMOCAP and Emo-DB in Figure~\ref{w}, where the results are obtained from test samples of the different speakers in IEMOCAP and Emo-DB.

From Figure~\ref{w}(a), it can be observed that when the DJDA model's training iterations progressively increase, the value of $w$ from 0.5 to over 0.9. This observation suggests that the contribution of marginal distribution adaptation of emotional features among different speakers in feature learning is weaken, and the conditional distribution adaptation plays a dominant role. While the results shown in Figure~\ref{w}(b) indicate a gradual transition from conditional distribution adaptation to joint distribution adaptation when dealing with speakers from Emo-DB.
These findings validate the effectiveness of the proposed DJDA in adaptively adjusting the contributions of MDA and CDA, indicating that our DJDA allows to more flexible handle unknown data distributions from different speakers.

\section{Conclusions}

In this paper, we proposed a Dynamic Joint Distribution Adaptation (DJDA) for speaker-independent SER. DJDA was designed to address multi-domain unsupervised domain adaptation, aiming to align the global feature distribution and local distribution of emotional features from different speakers more accurately. Experimental results on two publicly speech emotion datasets demonstrated the superior performance of the proposed DJDA. In future work, we will explore the dynamic joint distribution adaptation strategy when the emotion category feature spaces of the source and target domains are different.

\section{Acknowledgements}

This work was supported in part by the National Key R\&D Project under the Grant 2022YFC2405600, in part by the NSFC under the Grant U2003207 and 61921004, in part by the Jiangsu Frontier Technology Basic Research Project under the Grant BK20192004, in part by the YESS Program by CAST (Grant No. 2023QNRC001) and JSAST (Grant No. JSTJ-2023-XH033), in part by the ASFC under the Grant 2023Z071069003, in part by the China Postdoctoral Science Foundation under the Grant 2023M740600, and in part by the Jiangsu Province Excellent Postdoctoral Program.
\small
\bibliographystyle{IEEEbib}
\bibliography{mybib}

\end{document}